\documentclass[showpacs,aps,prb,amsmath,amssymb,superscriptaddress]{revtex4}
\usepackage{graphicx}
\usepackage{dcolumn}
\usepackage{bm}
\begin{document}

\baselineskip=0.55cm
\renewcommand{\thefigure}{\arabic{figure}}
\title{Conductance of a disordered graphene superlattice}

\author{N. Abedpour}
\affiliation{Department of Physics, Sharif University of Technology,
11365-9161, Tehran, Iran}
\author{Ayoub Esmailpour }
\affiliation { Department of physics, Shahid Rajaei University , Lavizan,
Tehran 16788, Iran}
\affiliation{School of Physics, Institute for research in fundamental
sciences, IPM 19395-5531 Tehran, Iran }
\author{Reza Asgari}
\affiliation{School of Physics, Institute for research in fundamental
sciences, IPM 19395-5531 Tehran, Iran }
\author{M. Reza Rahimi Tabar}
\affiliation{Department of Physics, Sharif University of Technology,
11365-9161, Tehran, Iran}
\affiliation{Institute of Physics, Carl von Ossietzky University, D-26111
Oldenburg, Germany}

\begin{abstract}

We study the conductance of disordered graphene
superlattices with short-range structural correlations. The system
consists of electron- and hole-doped graphenes of various
thicknesses, which fluctuate randomly around their mean value. The
effect of the randomness on the probability of transmission through
the system of various sizes is studied. We show that in a disordered
superlattice the quasiparticle that approaches the barrier interface
almost perpendicularly transmits through the system. The
conductivity of the finite-size system is computed and shown that the conductance
vanishes when the sample size becomes very large, whereas for some
specific structures the conductance tends to a nonzero value in the
thermodynamics limit.

\end{abstract}

\maketitle

\section{Introduction}

Graphene, a single atomic layer of graphite, has been successfully produced in
experiment~\cite{novoselov}, which has resulted in intensive investigations on
graphene-based structures, due to the fundamental physics interests that is
involved and the promising applications~\cite{geim}. There are significant
current efforts devoted to growing graphene epitaxially~\cite{berger} by
thermal decomposition of silicon carbide (SiC), or by vapor deposition of
hydrocarbons on catalytic metallic surfaces, which could later be etched away,
leaving graphene on an insulating substrate. The low energy quasiparticle
excitations in graphene are linearly dispersing, and are described by Dirac
cones at the edges of the first Brillouin zone. The linear energy-momentum
dispersion has been confirmed by recent observations~\cite{novoselov1}.
The slope of the linear relation corresponds the Fermi velocity of chiral
Dirac electrons in graphene, which plays an essential role in the Landau-Fermi
liquid theory~\cite{polini} and has a direct connection to the experimental
measurement.

There are some unusual features of graphene, such as the effects of
electron-electron interactions on the ground-state properties~\cite{yafis},
anomalous tunneling effect described by the Klein tunneling, the tunneling
through a p-n junction\cite{klein,katsnelson1} that follows from chiral band
states, and the energy-momentum linear dispersion relation. The Klein tunneling
predicts that the chiral massless carrier can pass through a high
electrostatic potential barrier with probability one, regardless of
the height and width of the barrier at normal incidence, which is in
contrast with the conventional nonrelativistic massive carrier
tunneling where the transmission probability decays exponentially
with the increasing of the barrier hight and would depend on the
profile of the barrier.~\cite{greiner,su,dombey,krekora}

An exciting experimental development is the ability to apply an electric field
effect or submicron gate voltage, in order to illustrate graphene p-n
junctions.~\cite{williams} By applying an external gate voltage, the
system can be switched from the n-type to the p-type carriers, thereby
controlling the electronic properties that give rise to graphene-based
nanodevices. Recently, strong evidence for Klein tunneling across
potential steps which is steep enough in graphene has been
experimentally observed.~\cite{stander}

Clean graphene junctions were predicted to display a number of
fascinating physical phenomena, even in the absence of electron-electron
interactions.~\cite{cheianov} Interestingly, the Veselago lensing of
electric current by a single p-n junction in clean
graphene~\cite{cheianov1} and the Andreev reflection, and the electron to
hole conversion at the interface at normal
incidence~\cite{beenakker}, have all been predicted theoretically. Such
phenomena are predicted to change both quantitatively and
qualitatively when disorder is included in the model. For instance,
inhomogeneous graphene p-n junction systems were studied using
the Thomas-Fermi approximation, including disorder effects, by Fogler
and collaborators~\cite{fogler}. They showed that junction
resistance is dominated by either ballistic or diffusive
contributions, depending on the density of charged impurity and gradient
of the carrier density.

In the semiconductor context there are basically a large number of
works on the tunneling, which have resulted in the "obvious" declaration that
the electronic properties of semiconductor superlattice are different from
those calculated in a single-barrier junction. Moreover, the
electronic properties of semiconductor superlattices in the presence
of disorder have been studied by several
groups~\cite{diez1,adame,diez2,bellani,esmailpour}. Importantly, all the
electronic states are localized in the thermodynamic limit for a
semiconductor superlattice in the presence of white-noise
disorder.~\cite{diez2}

Graphene superlattices, on the other hand, may be fabricated by adsorbing
adatoms on graphene surface through similar techniques, by positioning and
aligning impurities with scanning tunneling microscopy~\cite{eigler} or
by applying a local top gate voltage to graphene.~\cite{huard}
Recently, a periodic pattern in the scanning tunneling microscope image has
been demonstrated on a graphene on top of a metallic substrate.~\cite{marchini}
The transition of hitting massless particles in graphene-based
superlattice structure (GSLs) was first studied by
Bai and Zhang.~\cite{bai} They showed that the conductivity of
the GSLs depends on the superlattice structural parameters. Furthermore,
the superlattice structure of graphene nanoribbons
has been recently studied by using first-principles density
functional theory calculations.~\cite{sevincli} These calculations
showed that the magnetic ground state of the constituent ribbons, the
symmetry of the junction, and their functionalization by adatoms represent
structural parameters to the electronic and magnetic
properties of such structures.
Recently, novel physical properties of GSLs with one-dimensional (1D)
Kronig-Penney type and 2D muffin-tin type potentials
were also studied.~\cite{cheol} The results showed that a periodic potential
applied by suitably patterned modifications leads to further charge carrier
behavior. The propagation of charge carriers through such a superlattice is
highly anisotropic, and in extreme cases results in group velocities that are
reduced to zero in one direction but are unchanged in the other direction.
Moreover, they showed that the density and type of carrier are
extremely sensitive to the applied potential.

It would, therefore, be worthwhile to investigate how the conductance of
graphene superlattice junctions are affected by structural white noise, and
compare the conductances with those calculated for disordered
semiconductor superlattice. Due to the conservation of pseudospins in
graphene, backscattering process is suppressed at normal incidence,
which makes the disordered regions transparent.~\cite{beenakker}

The purpose of this paper is to study the electronic behavior of
graphene superlattices p-n junctions by using the transfer-matrix
method. The system that we study consists of a sequence of electron-doped
graphene as wells, and hole-doped graphene as barriers. We study the effect of
the disorder imposed on the size of the barriers in the transmission
probability, $T$, through the system as a function of the system size
(number of the barriers), together with various incident angles. The
dc conductance of the finite-size system takes on a nonzero value of the
transmission in some special configurations. Using the finite-size
scaling of transmission, we show that the conductance, in the thermodynamic
limit, tends to a finite constant for spacial cases.

The rest of this paper is organized as follows. In Sec. II we
introduce the models and derive the related transfer matrix. We also
explain how we calculate the transmission probability and the dc conductivity.
Section III contains our numerical calculations. We conclude
in Sec. IV with a brief summary.

\section{Model and Theory}

Consider a system of superlattice p-n junctions in the independent carriers
model at zero temperature, and in the absence of carrier-phonon and spin-orbit
interactions. The low-energy massless Dirac-band Hamiltonian of
graphene in the continuum model can be written as\cite{slon,haldane}
\begin{center}
${\cal H}_0=\hbar v \tau \left(\sigma_1 \, k_1 + \sigma_2 \, k_2
\right)$
\end{center}
where $\tau=\pm 1$ for the inequivalent $K$ and $K'$ valleys at which $\pi$
and $\pi^*$ bands touch, $k_i$ is an envelope function momentum operator, $v$
is the Fermi velocity, and $\sigma_i$ is a Pauli matrix that acts on the
sublattice pseudospin degree of freedom. The total Hamiltonian of a massless
carrier in a special geometry is written as, ${\cal H}={\cal H}_0+V(x)$ where
$V(x)$ is the graphene-based superlattice potential which is modeled as
described below.

\subsection{Superlattice Model}

We consider superlattice p-n junctions in a graphene-based
structure. The system consists of two kinds of graphene with
different potentials, the first being an electron-doped graphene with
thickness $d_W$, while the second is a hole-doped part with thickness
$d_B$, standing alternately. The potential for the electron- and hole-doped
graphene are $V_0$ and zero, respectively. The energy of the incident
particle is $E_0=2\pi\hbar v/\lambda$ with the wavelength $\lambda$
across the barriers, in such a way that the Fermi level lies in the
conduction band outside the barrier and the valence band inside it,
i.e., $(0<E_0<V_0)$, as shown in Fig. 1. The growth direction is
taken to be the $x$ axis which is designed as the superlattice axis.
In order to neglect the strip edges, we assume that the width of
the graphene strip is much larger than $d_B$. We set disorder situations
in which the value of $d_B$ fluctuates around its mean value,
given by $<d_B>=b$. In the model the fluctuations are given by,
$d_B|_i=b(1+\delta~\epsilon_i)$, where $\{\epsilon_i\}$ is a set
of uncorrelated random variables or white noise with the box
distribution, $-1\leq\epsilon_i\leq1$, and $i$ is the site index.
Here, the $\delta$ is the disorder strength.
\begin{figure}[ht]
\begin{center}
\includegraphics[width=0.7\linewidth]{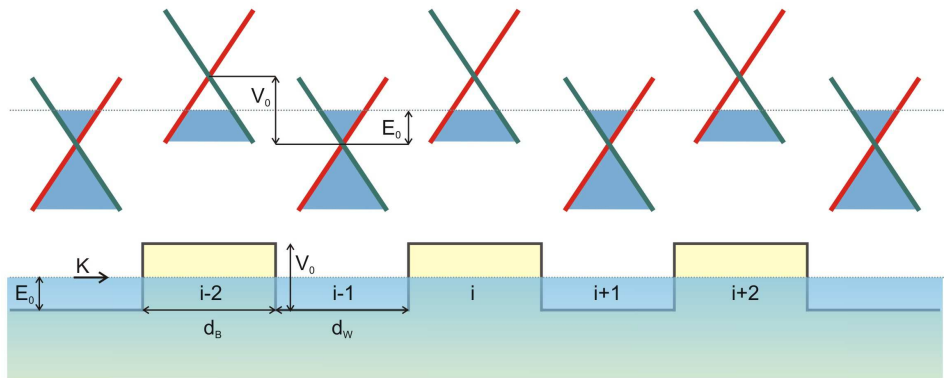}
\caption{Model of graphene superlattice p-n junctions.}\label{fig:sl}
\end{center}
\end{figure}

We consider graphene-based superlattice potential in a simple model as
\begin{eqnarray}
V(x) =
\left\{%
\begin{array}{ll}
V_0 \hskip 4.0cm  & \hbox{ $ {\rm if} \hskip 0.3cm |x-x_{2i}| <
\frac{d_B|_{i}}{2}$}\\
0 \hskip 4.0cm & \hbox{${\rm otherwise}$}~, \\
\end{array}%
\right.
\end{eqnarray}
where $x_{2i}$ is the position of barriers' center. The model
is similar to the potential of semiconductor superlattices that
has been used by other groups.~\cite{esmailpour}

\subsection{DC conductivity}

Let us now consider the case in which the incident massless electron in the GSLs
propagates at angle $\phi$ along the $x$ axis (see Fig. 1) and, therefore, the
Dirac spinor components, $\psi_1$ and $\psi_2$, which are the solutions to the
Dirac Hamiltonian, can be expressed~\cite{bai} as:
\begin{eqnarray}
\psi_1(x,y)&=&(a_i e^{i K_{i} x}+b_i e^{-i K_{i} x})e^{i k_{y}
y}\nonumber\\  \psi_2(x,y)&=&s_i(a_i e^{i K_{i} x+ i \phi_i}-b_i
e^{-i K_{i} x - i \phi_i})e^{i k_{y} y}~,
\end{eqnarray}
where
\begin{equation}
s_i=sgn(E_0-V(x)),  \hskip 0.8cm
k_{y}=\frac{E_0}{\hbar v} \sin(\phi)~,
\end{equation}
and
\begin{equation}
K_{i}=\left\{\begin{array}{cc}
          k_x=E_0 \cos(\phi)/\hbar v &  for \,\,\,\,\  well \\
          q_x=\sqrt{(E_0-V_0)^2/\hbar^2 v^2-k_y^2} & \,\,\ for
\,\,\,\,\  barrier
        \end{array}
        \right .
\end{equation}
In order to calculate the transmission coefficients, we use the
transfer-matrix method~\cite{katsnelson}. To this end, we apply
the continuity of the wave function at the boundaries, and construct the
transfer matrices as follows
\begin{eqnarray}\label{eq:t}
\left(
  \begin{array}{c}
    1 \\
    r \\
  \end{array}
\right)=\frac{1}{2\cos \phi}\left(
                             \begin{array}{cc}
                               e^{-i\phi}-e^{i\theta} &
e^{-i\phi}+e^{-i\theta} \\
                               e^{i\phi}+e^{i\theta} & e^{i\phi}-e^{-i\theta} \\
                             \end{array}
                           \right)P(2N)\times\\ \nonumber
                                            \left(
                                              \begin{array}{c}
                                                e^{i k_x
l_n}(e^{-i\theta}-e^{i\phi})/[2e^{i q_x l_n}\cos \theta] \\
                                                e^{i k_x
l_n}(e^{i\theta}+e^{i\phi})/[2e^{-i q_x l_n}\cos \theta] \\
                                              \end{array}
                                            \right) t_{2N}
\end{eqnarray}
where $r$ and $t_{2N}$ are the reflection and transmission coefficients of
the system that consists of $N$ barriers, and $p(2N)$ is the transfer matrix
given by
\begin{eqnarray}
P(2N)=\prod_{i=3}^{2N} P_{i,i-1}\\
\nonumber P_{i,i-1}=\left(
\begin{array}{cc}
M_{11} & M_{12} \\
M_{21} & M_{22} \\
\end{array}
\right)
\end{eqnarray}
where also $P_{i,i-1}$ is a transfer matrix from site $i$ to $i-1$ and
$M_{ij}$ are given by
\begin{eqnarray}
M_{11}&=&e^{i K_{i}l_{(i-1)}}e^{-i K_{(i-1)}l_{(i-1)}}[
e^{-i\varphi_{(i-1)}}- e^{i \varphi_i}]/2\cos(\varphi_{(i-1)})\\
\nonumber M_{12}&=&e^{-i K_i l_{(i-1)}}e^{-i K_{(i-1)}l_{(i-1)}}[
e^{-i\varphi_{(i-1)}}+ e^{-i \varphi_i}]/2\cos(\varphi_{(i-1)})\\ \nonumber
M_{21}&=&e^{i K_{i}l_{(i-1)}}e^{i K_{(i-1)}l_{(i-1)}}[
e^{i\varphi_{(i-1)}}+ e^{i \varphi_i}]/2\cos(\varphi_{(i-1)})\\
\nonumber
M_{22}&=&e^{-i K_{i}l_{(i-1)}}e^{i K_{(i-1)}l_{(i-1)}}[
e^{i\varphi_{(i-1)}}- e^{-i \varphi_i}]/2\cos(\varphi_{(i-1)})
\end{eqnarray}
where $l_i=\sum_{j=1}^{j=int[i/2]} d_B|_j +int[(i-1)/2]d_W$ is the
length of system at $i$-th boundary and moreover,
\begin{equation}
\varphi_i=\left\{\begin{array}{cc}
          \phi & for \,\,\,\,well \\
          \theta=\tan^{-1}(k_y/q_x) &\,\,\, for\,\,\,\,barrier
        \end{array}
        \right .
\end{equation}

It is evident that $T(E_0,\phi)=|t_{2N}|^2$, and that it can be calculated
from Eq. (\ref{eq:t}) for a given $N$. When the transmission
coefficients are calculated, the conductivity of system is computed
by means of the B\"uttiker formula,~\cite{datta} taking the
integral of $T(E_0,\phi)$ over the angle
\begin{equation}
G = G_0 \int^{\frac{\pi}{2}}_{-\frac{\pi}{2}}
T(E,\phi)\cos(\phi)d\phi
\label{eq:dc}
\end{equation}
where $G_0=e^2 m v w/\hbar^2$ with $w$ being the width of the graphene
strip along the $y$ direction.

\section{Results and Discussion}

Let us first calculate the
transmission probability and study the electronic properties of
disordered GSLs as a function of the strength of disorder introduced
in the system. We consider the width of barriers as a random
variable, so that the length of system in the numerical calculations
will be $L=N(b +d_W )$. In all the numerical calculations, we assumed
$b=<d_B>=50$ nm, while the wavelength of the incident particle is set by
$\lambda=50$ nm, or, equivalently, the energy of the carrier, $E_0=83$ meV.
In all of the calculations we used, $V_0=200$ meV, unless otherwise specified.
The number of realization of the random configurations is about 500.

Figure 2 shows the transmission probability, $T$, of the incident electrons
hitting a GSLs, as a function of the angle $\phi$ for several values
of the disorder strength, $\delta$. The number of the barriers in the
figure is, $N=100$, with $d_W=10$ nm. It is clear that, the
transmission decreases by increasing the disorder for all the angles, apart
from the strictly normal incidence case, $\phi=0$. This is
physically understandable due to the Klein tunneling process in graphene,
where the backscattering process is suppressed. Moreover, the
massless carriers with incident angle close to normal incidence can
survive in the presence of disorder, while the width of the angles around
the normal incidence decreases with increasing strength of the disorder as
well.

In order to understand the finite-size effect and the effect of the
width of the wells, the transmissions probability of a massless particle
through the system were calculated as a function of the incident angle. The
results are shown in Fig. 3 for several sizes at $\delta=0.1$. In Fig. 3(a)
we set $d_W=10$ nm. The transmission decreases with increasing system
size for all the angels, except again at $\phi=0$. This behavior is in
contrast with a clean GSLs result, where the number of the peaks
increases by increasing the number of the barriers.~\cite{bai} In
Fig. 3(b) the width of the wells is set to, $d_w=30$ nm. Two sharp peaks in the
transmission are obtained that disappear in Fig. 3(a).
Furthermore, in the case with $d_W=50$ nm there is only one angle,
$\phi\approx 60^\circ$, that the transmission survives in the presence of
the disorder, whereas under other angles the transmissions are suppressed.
Meanwhile, there is clearly a wide domain around $\phi=0$ for which the
transmission survives, and is larger than the one shown in
Fig. 3(b). The width of the domain decreases with increasing strength of the
disorder. It is worthwhile to note qualitatively that for
$\delta=0$, when we have the condition that,
$(q_x+k_x)\times(b+d_W)=2m\pi$ ( $m$ is an integer), the
transmission has finite values at angles different from $\phi=0$. This
is due to the resonance process in a system with $N$ barriers.

We also studied how disorder, introduced in the GSLs, affects the
conductivity of the system. Hence, we also calculated
numerically the dc conductivity by using Eq. (\ref{eq:dc}), with the
white-noise structural disorder imposed on the system. Figure 4 shows the dc
conductance of the GSLs as a function of $V_0$ for various strengths of the
disorder, $\delta$. As shown in the figure, the conductivity of the GSLs
decreases by increasing the strength of the disorder. However, the conductivity
approaches a finite value, i.e., the existence of a finite conductivity
in finite-size disordered GSLs should be expected. In general, the resonance
condition is given by a function that yields,
$f(q_x, d_W, d_B)=m\pi$. For instance, for the case $N=1$, the
condition yields, $q_x d_B=m\pi$, as a result of which $T(\phi)$
would be an oscillating function of $q_x$. Note that $q_x$ is
determined by $V_0$. Consequently, this leads to a finite dc conductivity which
is an oscillating function of $V_0$. The observation of conductance
oscillations in extremely narrow graphene hetrostructures has been observed
experimentally.~\cite{young}

In Fig. 5 the conductivity of the system is plotted as a
function of $V_0$, where the width of the wells is $d_w=10$ nm with
$\delta=0.1$. The conductivity decreases with increasing the size of system.
The inset in the figure shows the conductivity of a clean GSLs as a function
of $V_0$ for several system sizes. It indicates that the dc conductance of
clean superlattice behaves uniquely for different sizes, but in a disordered
GSLs it decreases by increasing the size of system, as shown in Fig. 5. At a
constant strength of the disorder, changing $d_W$ may also change the
conductivity, as depicted in Fig. 6. It demonstrates that the conductivity
varies periodically with increasing $d_W$. As a result, in the disordered GSLs,
the dc conductance of finite-size systems depends on the structural parameters,
especially $d_W$.

To compute all results that have been presented so far, we considered a
system of finite size. Next, we wish to calculate the finite-size scaling
of $G/G_0$. For this purpose, we calculated the conductivity as a function of
the system size. The results are summarized in Fig. 7. Importantly, the
conductivity vanishes by a simple power law, except for the case for which,
$\lambda=d_W=50$ nm. In general, for $\lambda=md_W$ the conductivity
approaches a finite value as $N$ becomes large.

In order to examine such results better, we also calculated the $G/G_0$ for a
case for which $\lambda=d_W=45$ nm. We found that the conductance tends to a
constant in the thermodynamic limit. the numerical data are fitted by using
\begin{equation}
\frac{G}{G_{0}} = g_{\infty} + \frac{\gamma}{L^\eta} \label{model}
\end{equation}
where $g_{\infty}$, $\gamma$, and $\eta$  are constants. $g_{\infty}$ is the
asymptotic value of $G/G_0$ in the thermodynamic limit, $N \to\infty$. As a
result, for the case $d_W=5$ nm we obtain $g_{\infty}=0$,
$\gamma\simeq 1.0$, and $ \eta\simeq 0.46$; for $d_W=10$ nm we obtained,
$g_{\infty}=0$, $\gamma\simeq 0.9$, and $\eta\simeq 0.42$, and
$g_{\infty}\simeq 0.14$, $\gamma\simeq 0.6$ and $\eta\simeq 0.2$ for $d_W=50$
nm. In all the case the regression was with $r^2=0.99$, indicating very
accurate fits. Note that $g_{\infty}$ is zero for small $d_W$, but tends to a
nonzero constant for $d_W=50$ nm.

\section{Conclusion}

We studied numerically the dc conductance of a discorded graphene superlattice
p-n junctions for various values of the strength of structural disorder imposed
on the material. It was shown that there exists a width around the normal
incidence angle for which the transmission becomes finite in the presence of
structural white-noise disorder. That is, the white-noise disorder gives rise
the largest number of the peaks in the transmission, suppressed in the
thermodynamic limit but quasiparticles which approach almost perpendicularly
to the barriers transmit through the material. We also calculated the
conductivity of a finite-size disordered system and showed that the
conductivity decreases by increasing the system size but that there are
cases for which the conductance approaches a nonzero value. This result
is in contrast with the case of a clean (ordered) GSLs.~\cite{bai}
Furthermore, the results of the finite-size scaling computations predict a zero
conductance for all the GSLs, except for some special $d_W$ values for which
$\lambda=m d_W$, where $m$ is an integer, in which case the conductance tends
to a nonzero constant in the thermodynamic limit.

Apparently, such a feature is independent of the value of $b=<d_B>$.
Consequently, we predict a finite conductivity for a disordered GSLs
when the wavelength of incident particle is equal to $m d_W$. These
results are in complete contrast with those calculated for
disordered semiconductor superlattice which become
insulator.~\cite{diez1, diez2, bellani}  Our finding for the dc
conductance of the GSLs should be important to the design of electronic
nano-devices based on graphene superlattices. It would probably worthwhile to
extend the present work to the case in which a correlated noise is used. In
this case one must replace the white-noise with a proper short- or
long-range correlated noise.

\begin{acknowledgments}

R. A. would like to thank the International Center for Theoretical Physics,
Trieste for its hospitality during the period when part of this work
was carried out. A. E and N. A are supported by the IPM grant.
\end{acknowledgments}

\newpage

\begin{figure}[ht]
\begin{center}
\includegraphics[width=3.1cm]{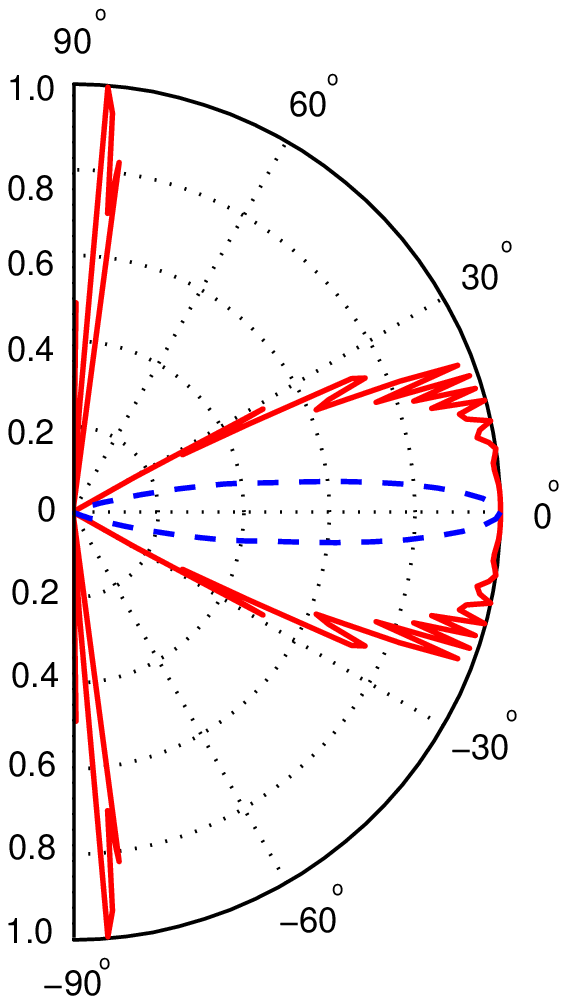}
\includegraphics[width=6.8cm]{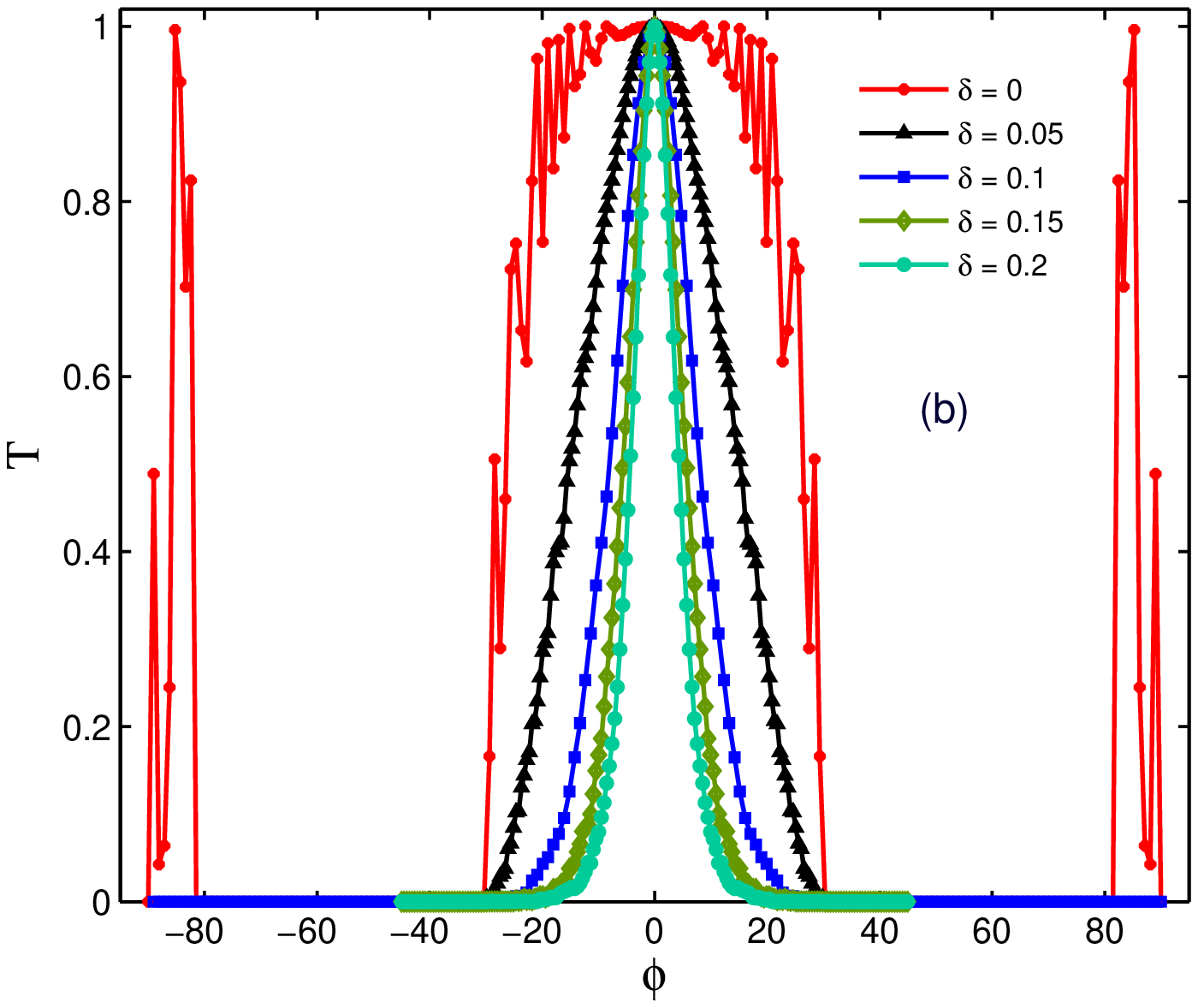}
\caption{(Color online) Transmission probability $T$ of electrons through the
system as a function of the incident angle for several disorder
strengths. (a): $\delta=0$ and $0.1$,  and (b): $\delta=0.0, 0.05, 0.1,
0.15$ and $0.2$ for $N=100$ and $d_W=10$nm.}
\end{center}
\end{figure}

\begin{figure}[ht]
\begin{center}
\includegraphics[width=6.2 cm]{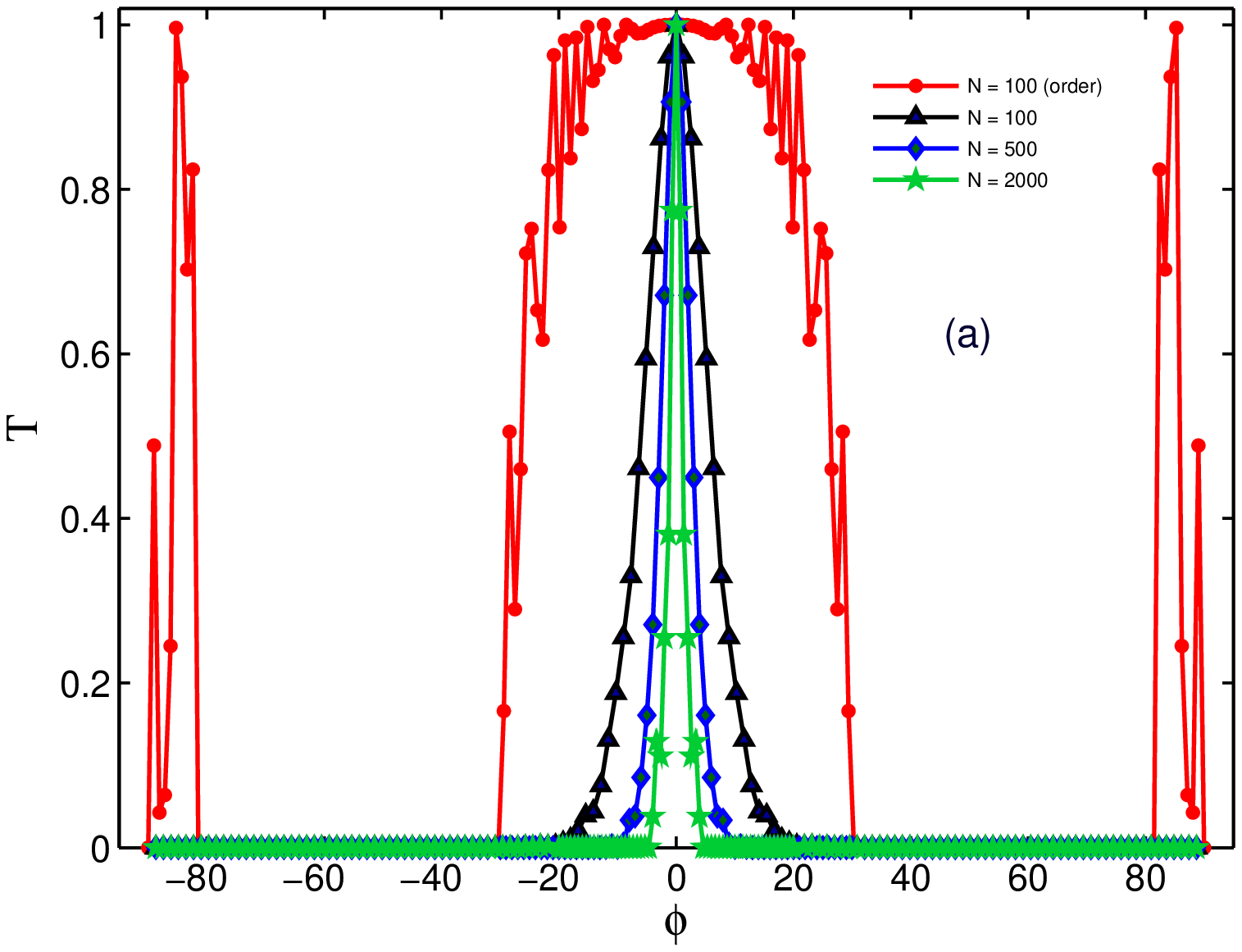}
\includegraphics[width=6 cm]{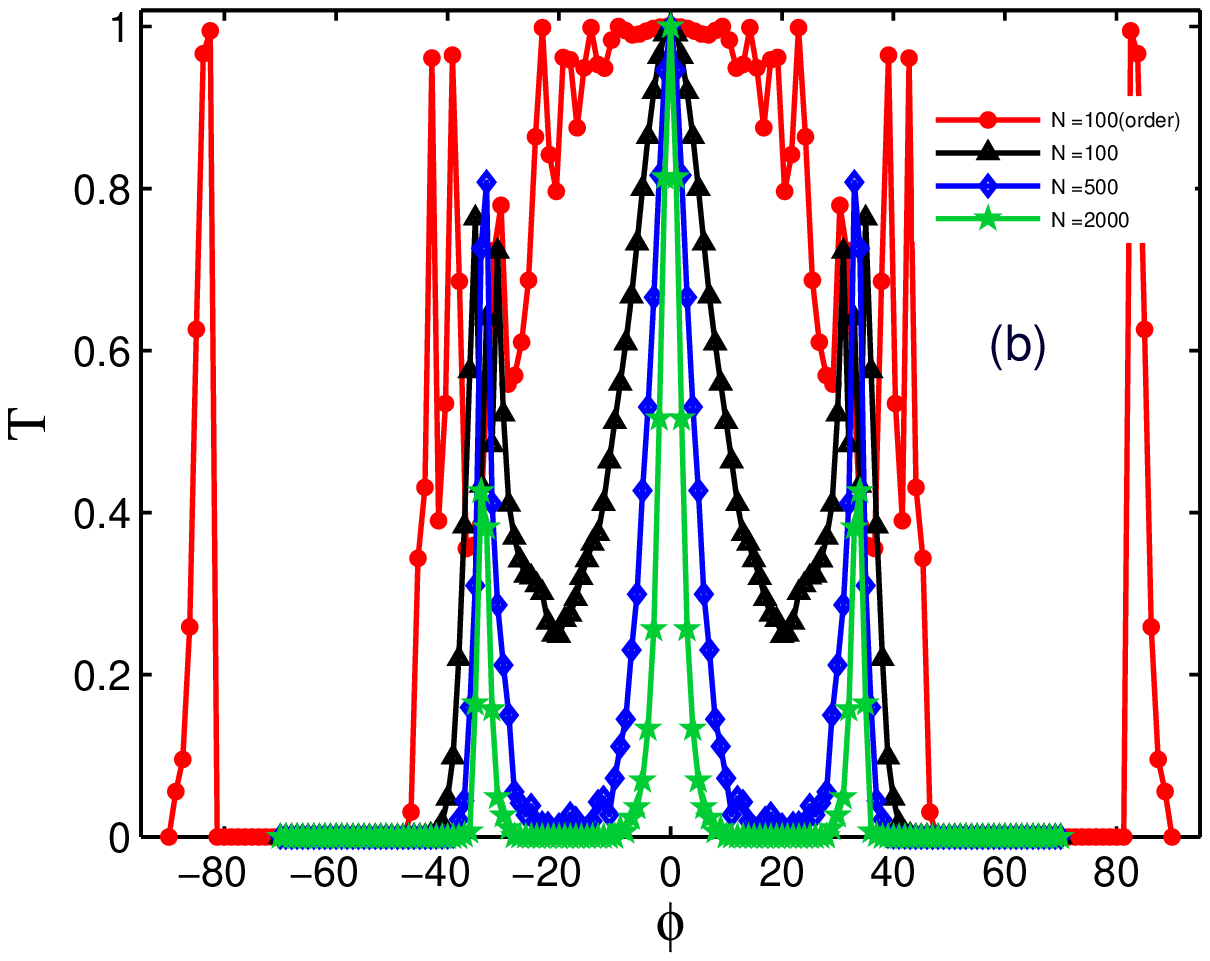}
\includegraphics[width=6.2 cm]{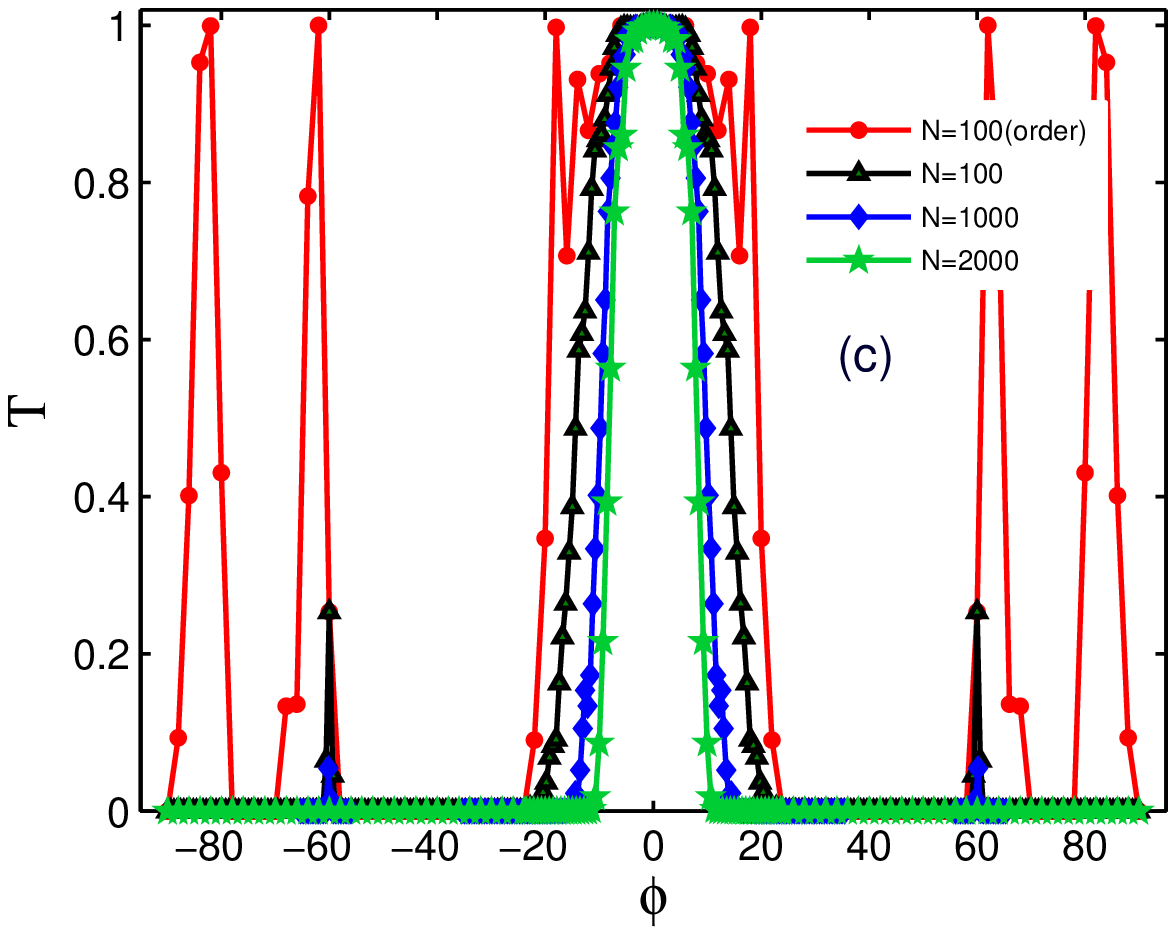}
\caption{(Color online) Transmission probability $T$ of the massless
carriers through the system as a function of the incident angle, the system
size and $\delta=0.1$ for $d_W=10$ nm. (a) $d_W=30$ nm. (b) $d_W = 50$
nm.}
\end{center}
\end{figure}

\begin{figure}[ht]
\begin{center}
\includegraphics[width=8 cm]{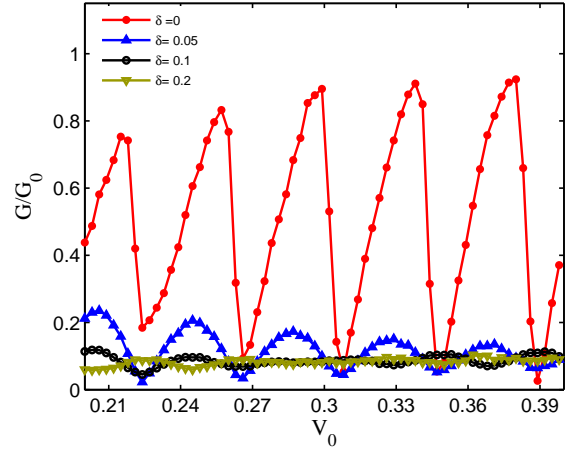}
\caption{(Color online) DC conductivity as a function of the barrier
potential, $V_0$ (in units of meV) for various strengths of the
disorder and $N=100$ and $d_W=10$ nm.}
\end{center}
\end{figure}

\begin{figure}[ht]
\begin{center}
\includegraphics[width=8 cm]{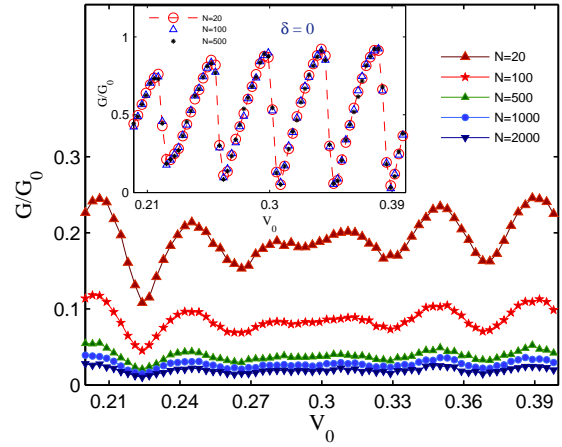}
\caption{(Color online) DC conductivity as a function of the barrier
potential $V_0$ (in units of meV) and system size for $\delta=0.1$ and
$d_W=10$ nm. Inset shows the same, but for clean (ordered) GSLs.}
\end{center}
\end{figure}

\begin{figure}[ht]
\begin{center}
\includegraphics[width=8 cm]{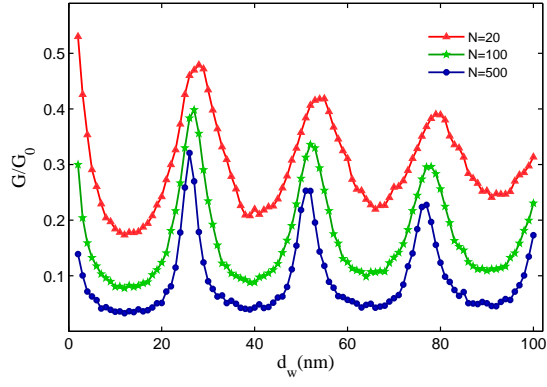}
\caption{(Color online) DC conductivity as a function of $d_W$ and
system sizes for $\delta=0.1$ and $V_0=300$ meV.}
\end{center}
\end{figure}

\begin{figure}[ht]
\begin{center}
\includegraphics[width=8 cm]{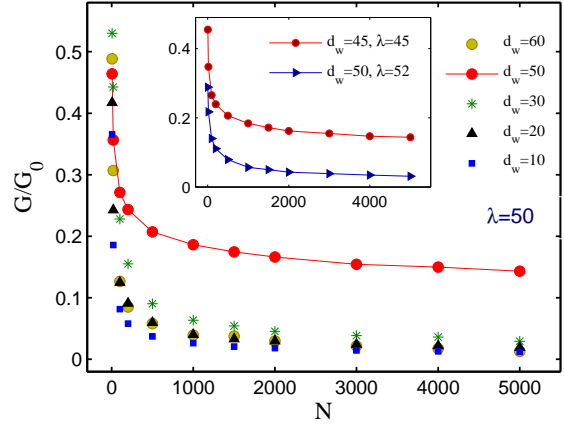}
\caption{(Color online) Finite-size scaling of the dc conductivity as a
function of size $N$ for the disorder strength $\delta=0.1$ and various values
of $d_W$ and $\lambda$ (in the inset).}
\end{center}
\end{figure}

\end{document}